# Could the EPR correlation be in superconducting structures? Possibility of experimental verification.

V.V. Aristov and A.V. Nikulov

Institute of Microelectronics Technology, Russian Academy of Sciences, 142432 Chernogolovka, Moscow District, Russia. E-mail addresses: nikulov@ipmt-hpm.ac.ru.

## ABSTRACT

In order to have a chance to make a real quantum computer it is important to find the entanglement phenomenon on mesoscopic level since technology can not be able in the visible future to work on atomic level. It is known that the entanglement (the EPR correlation) violates the principle of local realism. Existence of EPR correlation on a level higher than atomic one violates both local and macroscopic realism. In order to clear possibility of this violation the fundamental difference between applications of the principles of quantum mechanics on atomic and higher level is considered. Mysterious nature of some macroscopic quantum phenomena is accentuated. Experimental verification of the entanglement possibility in superconductor structure is proposed.

## 1. INTRODUCTION

Anthony Leggett writes in[1]: "*Quantum mechanics has been enormously successful in describing nature at the atomic level and most physicists believe that it is in principle the "whole truth" about the world even at the everyday level*". On the other hand Richard Feynman, in his book *The Character of Physical Law*, declared that "*nobody understand quantum mechanics*". According to these opinions of the two Nobel Prize Winners, quantum mechanics, for most physicists, is matter rather of belief than understanding. Like situation was in the end of 19 century when "*because of the trust in great successes achieved with thermodynamics the Carnot's principle was elevated to the class of absolute, precise dogma*" (the quotation from[2]. Because of this belief in the Carnot's principle, which "*we call since Clausius time the second law of thermodynamics*"[2], most 19th century physicists rejected the atomic-kinetic theory of the heat proposed by Maxwell and Boltzmann.

Since the quantum mechanics is matter rather of belief than understanding for most 20th century physicists they disregarded the philosophical controversy which was between founders of quantum mechanics. There is observed a scholastic point of view. All modern physicists studied quantum physics in school and heard that its principles are paradoxical. This paradoxicality has became habitual and most physicist do not observe it. Richard Feynman said that habit may seem understanding. This habit results to an illusion of understanding of quantum physics although some experts understand that "*in contrast to the theories of relativity, quantum mechanics is not yet based on a generally accepted conceptual foundation*"[3]. This illusion may be most dangerous now when, as some experts remark[1], the problem discussed by the founders of quantum mechanics can not be considered as purely philosophical but have a direct relation to modern investigation. Moreover these problems have now a practical importance in view of the idea the quantum computation[4,5,6].

The quantum computation is one of the most intriguing ideas of the last years and a practical realization of a real quantum computer is one of the most grandiose problems. This idea attract attention of many scientists. First proposals of realization of this idea were made for atomic-level systems[7-10]. But there is few chance that technology will be able to work on the atomic level in the near future. Therefore the only chance to make the quantum computer in the near future should be connected with the level higher than atomic one, the level is mastered already by the modern technology. Superconductivity is attractive for a realization of the idea of quantum computer since it is one of the macroscopic quantum phenomena.

The idea of the quantum computation is based on the very paradoxical quantum phenomenon called entanglement or Einstein - Podolsky - Rosen correlation. Nobody today understands what is the essence of entanglement. But it does not mean that we should not try to understand what is the essence of entanglement. Thanks to a reflection about this problem we can understand what can not be entanglement. Even suppositions proposed here may be useful. For example if the authors[11-13] proposing to achieve the entanglement in superconductor mesoscopic structures with help of classical interactions and authors[14-16] realizing these proposals knew that Erwin Schrodinger said about entanglement of our knowledge[17] it may be they did not propose this. It is obvious that no classical interaction can entangle our knowledge.

The paradoxical nature of the entanglement is exposed in the Einstein- Podolsky- Rosen paradox. The long history of this paradox demonstrates clearly that if we have studied anything it does not mean that we understand this. Nevertheless same physicists ignore the paradoxical nature of the EPR correlation and are sure that it is enough to write a Schrodinger equation in order to obtain the entanglement. Some physicists think that the observations of the Bell's correlation have shown that Einstein did not understand quantum physics. Indeed, Einstein did not studied quantum physics in school and they studied. The scholastic point of view prevents to realize that the violation of the principle of local realism is very great paradox. Unfortunately this point of view predominates now. Only therefore the proposals[11-13] to achieve the pure non-classical the EPR correlation with help of classical interactions could be published.

It should be obvious for any physicists that no classical interaction can contradict to the principle of the local realism. Moreover, no all quantum principles contradict to the local realism. It is not follow from the existence of the macroscopic quantum phenomena, such as superconductivity, that the principle of local realism should be violated on the macroscopic level. Violation of the local realism on this level means simultaneously violation of the macroscopic realism[18]. A number of experiments[19-26] have confirmed the quantum mechanical prediction contrary to the local realism but no reliable experimental evidence of violation of the principle of macroscopic realism has been obtained for the present. Moreover it is not enough clear where one ought search of this violation.

In order the search of the entanglement in superconductor could be intelligent the known experimental results obtained at investigation of superconductivity as quantum macroscopic phenomenon ought be considered again. We bring to light in the Chapter 2 mysterious nature of well known macroscopic quantum phenomena which is disregarded by most physicists because of the predominated scholastic point of view. It is important to realize the fundamental differences between application of paradoxical principle of quantum mechanics on atomic and higher levels. We considered some of these differences in the Chapter 3. The contradiction between principles of quantum mechanics and macroscopic realism is considered in the Chapter 4. Proposals to verify the possibility of entanglement in superconductor structure are expounded in the Chapter 5. The EPR paradox[27] and the Bell's experiments[19-26] testify that the entanglement is a combination of a law of conservation and quantum superposition. Our proposals are based on the law of conservation of momentum circulation of superconducting pairs in loops.

## 2. SUPERDCONDUCTIVITY IS MACROSCOPIC QUANTUM PHENOMENON

The paradoxical principles of quantum mechanics were pressed by experimental results obtained on atomic level. Founders of quantum mechanics applied these principles first of all to atomic physics. For example, Niels Bohr wrote in the answer[28] on the critique of his interpretation of quantum mechanics by A. Einstein, B. Podolsky, and N. Rosen[27]: "The trend of their argumentation, however, does not seem to me adequately to meet the actual situation with which we are faced in atomic physics". Just in the year of this discussion between founders of quantum physics the first phenomenological theory of the Meissner effect has been put forth by the London brothers[29]. We know now that the Meissner effect is experimental evidence of macroscopic quantum phenomenon. But the London's theory is based on a classical consideration.

**2.1 The London's theory**
For an outline the theory[29], one ought to start with the case of infinite electric conductivity. There is a relation $E = (\lambda_L^2/\mu_0)dj/dt$ between an electric $E$ and a current density $j = n_s qv$ in an ideal conductor with a density $n_s$ of a charge $q$ carriers since only electric force $qE$ acts on a particle with a mass $m$ without a dissipation force and therefore $mdv/dt = qE$. The quantity $\lambda_L = (\mu_0 m/n_s q^2)^{1/2}$ is represents a length of order 10 nm for a typical density of electrons $q = e$ in a metal. Inserting the Maxwell equation curl $E = -dB/dt = -\mu_0 dH/dt$ into the relation between $E$ and $dj/dt$, we have $\lambda_L^2$ curl $dj/dt + dH/dt = 0$. From the other Maxwell equation curl $H = j$ we then obtain $\lambda_L^2$ curl curl $dH/dt + dH/dt = 0$ or

$$\lambda_L^2 \nabla^2 \, dH/dt = dH/dt \qquad (1)$$

This equation has the solution $dH/dt = dH(0)/dt \exp(-x/\lambda_L)$ considering an ideal conductor filling the half-space with $x > 0$ and letting the coordinate x run from the surface (at $x = 0$) into the interior of the ideal conductor. According to this solution magnetic field H can not change in the interior of ideal the ideal conductor, $dH/dt = 0$ at $x \gg \lambda_L$.

But the solution of the equation (1) can not explain the expulsion of magnetic flux from the interior of a superconductor observed at the Meissner effect. There is fundamental difference between infinite electric conductivity, as a classical phenomenon, and superconductivity, as macroscopic quantum phenomenon. But the London brothers did not used principles of quantum mechanics, although these principles were already formulated in 1935. In order to explain the Meissner effect have removed the time derivation in the equation (1) and thereby have postulated the new equation

$$\lambda_L^2 \nabla^2 H = H \qquad (2)$$

yielding the solution $H = H(0) \exp(-x/\lambda_L)$ for the same geometrical conditions as before. This solution indicates that the magnetic field is exponentially screened from the interior of a superconductor, the screening taking place within a surface layer of thickness $\lambda_L$. The quantity $\lambda_L$ is generally referred to as the London penetration depth.

**2.2 The Meissner effect is consequence of the Bohr's quantization**
Meissner and Ochsenfeld observed in 1933 [30] that a superconductor, placed in a weak magnetic field, completely expels the field from the superconducting materials except for a thin layer at the surface. Bohr postulated the quantization $m_e v_e r_n = n\hbar$ of the moment $mv_e r_n$ of electron velocity $mv_e$ on atomic orbit in 1913. F. London assumed in 1948 [31] that the strangeness of superconductor electrodynamics may be connected with a long-range order of the momentum vector $p = mv + qA$ and introduced in 1950 [32] the concept of the fluxoid $\Phi l = \Phi + (m/q) \oint v_s dl = q^{-1} \oint p dl$. Permitted values of this quantity are $\Phi l = n(2\pi\hbar/q) = n\Phi_0$ because of the Bohr quantuzation $\oint p dl = n2\pi\hbar$, where n is an integer number and $\Phi_0 = 2\pi\hbar/q$ is the flux quantum. Macroscopic quantum mechanism of superconductivity was mentioned in the book[33] published in 1952.

The quantization of the fluxoid can also be deduced from the requirement that the complex wave function $\Psi = |\Psi| \exp i\varphi$, used in the Ginzburg-Landau theory[34], must be single-valued at any point in the superconductor. Therefore, the phase $\varphi$ of this wave function must change by integral multiples of following a complete turn along the path of integration, yielding the same Bohr quantum condition $\oint \nabla\varphi dl = n2\pi$ since $\nabla\varphi = p/\hbar$ in quantum physics. The Meissner effect has simple, mathematical explanation when the long-range phase $\varphi$ coherence of the wave function $\Psi(r) = |\Psi|exp(i\varphi)$ is used. It is well known that an integral along any closed path *l* of gradient of a function equals zero $\oint dl \nabla\varphi = 0$ if this function does not have singularities inside the closed path *l*. It is important that the gradient of phase of the wave function is proportional to the momentum of superconducting pairs $\hbar\nabla\varphi = p = mv + 2eA$. Where *A* is the vector potential; *v* is the velocity of the superconducting pairs. Therefore when $\oint dl \nabla\varphi = 0$ then the magnetic flux $\Phi = \oint dl A$ contained within the closed path *l* should be equal zero $\Phi = \oint dl A = (\hbar/2e)\oint dl \nabla\varphi - (m/2e)\oint dl v = 0$ if the velocity circulation equals zero $\oint dl v = 0$.

**2.3 Mystique of the Meissner effect**
Numerous experimental results corroborate the fundamental difference between superconductor and ideal conductor. Nobody doubts now that superconductivity is macroscopic quantum phenomenon. But only few experts[35] realize that the expulsion of magnetic flux from the interior of a superconductor is mysterious phenomenon. The contradiction to the law of momentum conservation is obvious from the well known difference between superconductivity and infinite electric conductivity. Electrodynamics properties of an ideal conductor corresponds to the law of momentum conservation. Therefore the difference of the behaviour of superconductor from the one of ideal conductor manifesting in the Meissner should raise a question about the momentum conservation. When the London brothers explained the Meissner effect they did not come into conflict with this law until the removing of the time derivation in the equation (1) obtained from the classical laws. This removing contradicts to the bases of classical physics but is grounded by the Bohr's quantization

$$M_l = \oint dl p = \oint dl(mv_s + qA) = m\oint dl v_s + q\Phi = n2\pi\hbar \qquad (3)$$

There should be the difference $n2\pi\hbar - q\Phi = 2\pi\hbar(n - \Phi/\Phi_0)$ between the equilibrium values of the momentum circulation at $M_l = n2\pi\hbar$ and without $M_l = q\Phi$ the Bohr's quantization. The equilibrium average velocity of pair of electrons with the charge $q = 2e$ equals zero $v = 0$ in the normal state and its momentum circulation along a closed path l equals $M_l = q\Phi$. Where $\Phi = BS$, B is the magnetic induction induced by an external magnet, S is the inside l. The magnetic flux $\Phi$ is expulsed from the interior of a superconductor at the transition into superconducting state in accordance with the Meissner effect. Consequently the momentum circulation along any closed path l the interior of a superconductor changes from $M_l = q\Phi$ to $M_l = 0$ at the transformation of pair of electrons into the superconducting pairs. The momentum circulation equals zero $M_l = 0$ along any closed path l both in the interior of a superconductor and in the surface layer of thickness $\lambda_L$. But in contrast to the interior of a superconductor both the velocity circulation along l and the magnetic flux inside l are not equal zero when l is taken along superconductor surface.

There is the relation $M_l = m2\pi r v_s + q\Phi = 0$ for any circular path with a radius $r < r_{cyl}$ having common axis with long superconductor cylinder with the radius $r_{cyl}$ placed in a parallel magnetic field $H_{ext}$. This relation is equivalent to the relations $\lambda_L^2 2\pi r \mu_0 j_s + \Phi = -\lambda_L^2 2\pi r \mu_0 dH/dr + \Phi = -\lambda_L^2 d^2\Phi/dr^2 + \Phi = 0$ or $-\lambda_L^2 d^2H/dr^2 + H = 0$ since the superconducting current density $j_s = qn_s v_s$ and the London penetration depth $\lambda_L = (\mu_0 m/n_s q^2)^{1/2}$. Where $q = 2e$, $n_s$ and $v_s$ are the charge, the density and velocity of superconducting pair. These equations have the solutions $H = H_{ext} \exp(r - r_{cyl})/\lambda_L$ and $\Phi = \Phi_{ext} \exp(r - r_{cyl})/\lambda_L$, where $H_{ext}$ is the external magnetic field and $\Phi_{ext} = \lambda_L 2\pi r_{cyl} \mu_0 H_{ext}$ at $r_{cyl} \gg \lambda_L$. Experiments show that the Meissner effect is observed in macroscopic, enough large samples. The change of the magnetic flux $\lambda_L 2\pi r_{cyl} \mu_0 H_{ext} - \pi r_{cyl}^2 \mu_0 H_{ext} \approx -\pi r_{cyl}^2 \mu_0 H_{ext}$ can induce an enough high Faraday's electric field $2\pi r_{cyl} E = -d\Phi/dt \approx \pi r_{cyl}^2 \mu_0 H_{ext}/\tau_{Meis}$ along the cylinder circumference $2\pi r_{cyl}$ at a shot time $\tau_{Meis}$ of the magnetic flux expulsion. At the same time $\tau_{Meis}$ superconducting pairs accelerate along the cylinder circumference from $v_s = 0$ to $v_s = (q/m2\pi r_{cyl})\Phi_{ext} = (q\lambda_L/m)\mu_0 H_{ext}$. There is important to emphasize that pairs accelerate against the force $2eE$ of the Faraday's electric field.

It is strange, but only few physicists[35] take notice that superconducting pairs accelerate against the Faraday's electric field force at the Meissner effect. The author[35] thinks that a revision of the conventional understanding of superconductivity is required for an explanation of this mysterious phenomenon. But we do not think that the change of the canonical momentum without any force observed at the Meissner effect can be connected with superconductivity phenomenon. It is rather a mysterious quantum effect since the momentum change occurs because of the Bohr's quantization. There is violation of the force balance. The like violation takes place also at the observation of the persistent current, existing because of the Bohr's quantization, in mesoscopic loops with non-zero resistance. The conventional circular current in a loop with non-zero resistance $R_l > 0$ is maintained by the Faraday's voltage $R_l I = -d\Phi/dt$ and there is the force balance. The persistent current $I_p \neq 0$ is observed at a magnetic flux invariable in time $d\Phi/dt = 0$ and there is violation of the force balance when $I_p$ is observed in a loop with $R_l > 0$. This phenomenon observed in superconductor[36-38], normal metal[39-41] and semiconductor[42-44] loops is connected with the Bohr's quantization both in the normal state of superconductor[45] and in the normal metal[46] (or semiconductor). Thus, the problem with the force balance connected with the Bohr's quantization is observed not only in superconductor but also other mesoscopic structure. The Bohr's quantization is observed on atomic and higher level. In the atomic realm, charge particles do not change their state of motion in the absence of a force whereas we should assume such change on the higher level because of the observed violation of the force balance. This fact manifests a difference between application of paradoxical principles of quantum mechanics on atomic and higher levels.

### 3. FUNDAMENTAL DIFFERENCIES BETWEEN ATOMOIC AND HIGHER LEVLES

One of the three "axises" along which, according to A.J. Leggett[47], it is not unreasonable to seek evidence of a breakdown of the quantum mechanics scheme of the physical world is the collision of it with our immediate experience of the "everyday" world. The logic of the Copenhagen interpretation of the quantum mechanics is inseparably linked with some restrictions. For example, superposition of quantum states is possible only if noninvasive measurement is impossible. It is impossible logically to obtained at a single measurements two (or more) different values of a variable corresponding different quantum states. Therefore it is important to realize the difference of our possibilities and restriction of these possibilities on atomic and higher level. We draw attention in this Chapter to some fundamental differences between application of paradoxical principles of quantum mechanics on atomic and higher levels connected with this difference of possibilities.

### 3.1 Intrinsic breach of symmetry because of the Bohr's quantization

The quantization $m_e v_e r_n = p_e r_n = n\hbar$ of the moment $m v_e r_n$ of electron momentum $p_e = m v_e$ on atomic orbit postulated by Bohr has explained the possibility of stable electron orbits in atom. But there was a logical difficulty in the Bohr's model until electron considered as a particle having a velocity $v_e$ since it was impossible to answer on the question: "What direction has this velocity?" The uncertainty relation and the wave quantum mechanics have overcame this difficulty. According to the uncertainty relation $\Delta p_e \Delta l > \hbar$ there is no sense to say on a velocity $v_e = dl/dt$ since electron can not have any certain position $l$ on atom orbit. The Copenhagen interpretation of quantum mechanics has removed any question about a direction of some elementary processes. Albert Einstein considered it as weakness. He wrote many years ago: *The weakness of the theory lies ... in the fact, that it leaves time and direction of the elementary process to "chance"* (the citation from the paper[48]). But thanks to this "weakness" the Bohr's quantization does not violate symmetry on the atomic level. We can not say about a velocity direction of electron considered as wave on atom orbit. Therefore the discreteness of atomic levels does not result to an intrinsic breach of symmetry.

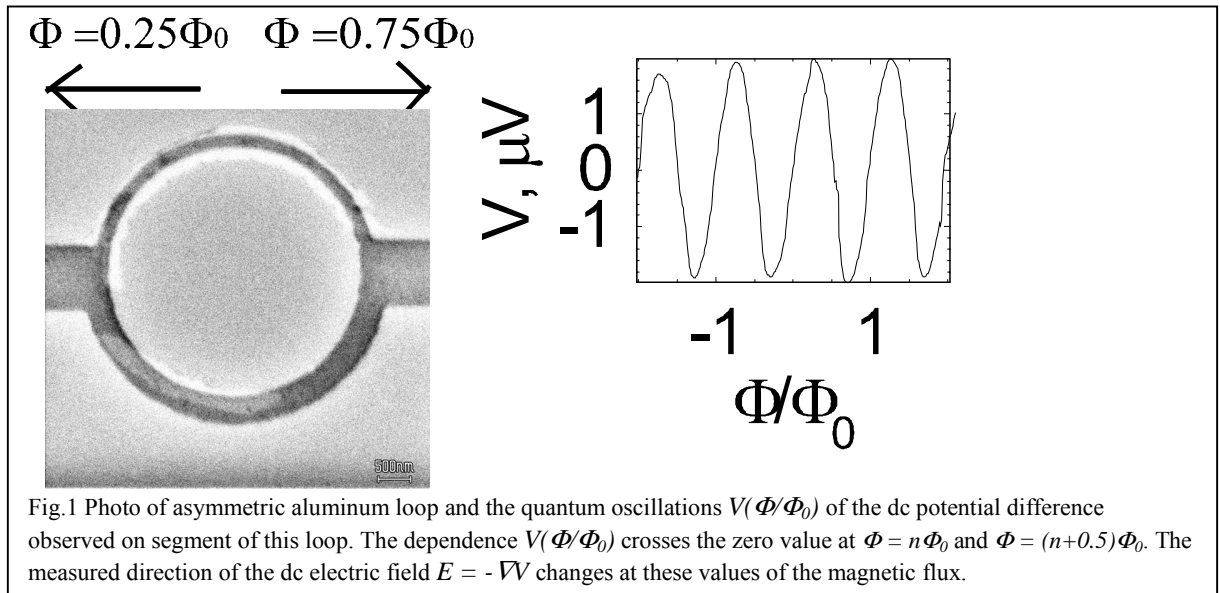

Fig.1 Photo of asymmetric aluminum loop and the quantum oscillations $V(\Phi/\Phi_0)$ of the dc potential difference observed on segment of this loop. The dependence $V(\Phi/\Phi_0)$ crosses the zero value at $\Phi = n\Phi_0$ and $\Phi = (n+0.5)\Phi_0$. The measured direction of the dc electric field $E = -\nabla V$ changes at these values of the magnetic flux.

The persistent current is observed because of the Bohr's quantization as well as stable electron orbit in atom. Therefore one could expect that the persistent current has not direction as well as the electron velocity on atom orbit. But experimental results force to come to a contrary conclusion. It was found that the dc potential difference $V(\Phi/\Phi_0)$ can be observed on segment of asymmetric superconducting loops with[49] and without[50,51] Josephson junctions. Sing and value of this dc voltage are periodical function $V(\Phi/\Phi_0)$ of magnetic flux $\Phi$ with period corresponding the flux quantum $\Phi_0 = \pi\hbar/e$ inside the loop[49-51]. This quantum oscillations can be observed without an evident power source in a temperature region near superconducting transition[50]. At a lower temperature they can be induced by an external ac current, for example $I_{ext}(t) = I_0 \sin(2\pi ft)$, when its amplitude $I_0$ exceeds a critical value $I_{0,c}$[51].

The observations[50,51] of the dc potential difference $V(\Phi/\Phi_0)$ on segment of superconducting loop, Fig.1, testify an analogy between the persistent current and a conventional circular current $I$ induced by the Faraday's voltage $R_l I = -$

$d\Phi/dt$ in a asymmetric loop l, like shown on Fig.1, with a resistance $R_l$, since the observed magnetic dependence $V(\Phi/\Phi_0)$ is like the one $I_p(\Phi/\Phi_0)$ of average equilibrium value of the persistent current. It is well known that the potential difference $V = (R_{ls} – R_l l_s/l)I = R_{asym}I$ should be observed of segment $l_s$, with a resistance $R_{ls}$ of conventional loop l with a resistance $R_l$ and a circular current I. Nobody can doubt that any conventional circular current has clockwise or contra-clockwise direction. This direction is determined by the direction of the Faraday's electric field $E_{Far} = -dA/dt$. On the other hand the current direction and asymmetry of the loop determine the direction of the potential electric field $E_p = -\nabla V$ observed on the loop segment. For example, when the circular current I in the loop shown on Fig.1 has clockwise direction then the potential electric field $E_p$ is directed from left to right. This violation of the symmetry between left and right direction is determined by the external vector factor - the Faraday's electric field. But the dc potential electric field $E_p = -\nabla V(\Phi/\Phi_0)$ is observed in a magnetic field invariable in time, i.e. without the Faraday's electric field $E_{Far} = -dA/dt = 0$. There is not also any other vector factor which could determined the observed direction $E_p = -\nabla V(\Phi/\Phi_0)$.

Therefore the observations[50,51] of the quantum oscillations of the dc voltage $V(\Phi/\Phi_0)$, Fig.1, is experimental evidence of intrinsic breach of symmetry. We should ask: "Why can the dc electric field $E_p = -\nabla V(\Phi/\Phi_0)$ have right direction (for example) at magnetic flux equal *0.25* of the flux quantum and left direction at *$\Phi = 0.75\Phi_0$*?" There can be only answer: "Since the persistent current has clockwise direction (for example) in the first case and contra-clockwise direction in the second one", since the quantum oscillations of the dc potential difference *$V(\Phi/\Phi_0)$* in magnetic field repeat the quantum oscillations $I_p(\Phi/\Phi_0)$ of the circular persistent current[36]. It seems self-evident that any direct current has a direction. But it is no so obvious for the persistent current existing because of the Bohr's quantization. There is an important difference between applications of quantum principles on atomic and higher levels: we can not make by hand asymmetric atom orbit, but we can make asymmetric mesoscopic loop. But we see from the experiment the intrinsic breach of symmetry because of the Bohr's quantization on the mesoscopic level. And we see that the direction returns in quantum physics on the mesoscopic level. The problem concerning philosophical foundation of quantum physics discussed by Albert Einstein and Niels Bohr remains topical problem for the present. Moreover this problem regarded as merely philosophical for many decades can be tested now in experiment. The experimental evidence of intrinsic breach of symmetry is fundamental result having significant consequences.

### 3.2 Intrinsic breach of symmetry of equilibrium motion and quantum limits to the second law
The experimental evidence of the intrinsic breach of symmetry forces to revise some bases of statistical physics and thermodynamics. The logic of statistical physics is based on some postulates seemed self-evident. Already founders of statistical physics, Maxwell and Boltzmann, postulated that average velocity of any equilibrium motion should be zero. This postulate was used 100 years ago as self-evident at the description of the Brownian motion by Einstein, Smoluchowski and others. They could not make doubt of the equal probability of motion in opposite directions since if anybody says that a right (or clockwise) direction has higher probability than opposite one he should explain why no left (or contra-clockwise) direction. This symmetry of Brownian motion seems self-evident because of the equality of directions in space. But this equality can be broken because of discreteness of permitted state spectrum. Opposite directions are not equivalent when a state with a velocity $v$ is permitted and the one with the velocity $–v$ is forbidden. The right (or clockwise) direction can have higher probability than opposite one in quantum system with discrete spectrum when, for example, the permitted state with lowest energy has right (or clockwise) direction of the velocity.

The intrinsic breach of symmetry of equilibrium motion observed on the mesoscopic level, i.e. the observations of a direct equilibrium motion is challenge to the second law of thermodynamics[52-54]. Richard Feynman[55] (and earlier by Smoluchowski[2]) have shown very well that the ratchet/pawl combination is not challenge to the second law since they can provide with a direct motion only at non-equilibrium conditions, when their temperature is lower than the temperature of molecules drawing the rotator. In this case, i.e. when molecules induce the direct, non-random, motion, an useful work can be extracted from heat. But it takes place in accordance with the Catnot's principle. The ratchet and pawl can not provide a direct motion under equilibrium conditions since they undergo the random equilibrium motion. In contrast to this the discreteness of the permitted state spectrum does not undergo the equilibrium motion but governs it. Therefore the persistent current, i.e. the direct circular equilibrium current having clockwise or contra-clockwise direction, is observed in mesoscopic quantum systems with discrete spectrum.

Most authors[56-58] interpret some phenomenon connected with the persistent current as a consequence of the ratchet effect or Brownian motors which provide a direct motion far from equilibrium. But this interpretation is not correct in principle. These authors do not take into account that the persistent current is already the direct motion observed under equilibrium conditions. They do not understand the fundamental difference between the discreteness of the spectrum and the ratchet/pawl combination.

According to numerous experimental results the persistent current is observed at non-zero resistance[36,38]. First evidence of it is the Little-Parks experiment[37] made in 1962. The resistance of a mesoscopic superconducting loop changes periodically with magnetic field $R_l(\Phi/\Phi_0)$, Fig.4, since the persistent current $I_p(\Phi/\Phi_0)$ changes periodically. Thus, the observation of the Little-Parks oscillations of the resistance is experimental evidence of the persistent current $I_p \neq 0$ at non-zero resistance $R_l > 0$. It is obvious already from the experimental results that the persistent current in this case is an ordered Brownian notion, i.e. a fluctuation phenomenon, since the Little-Parks oscillations $R_l(\Phi/\Phi_0)$ is observed only in the fluctuation region near the superconducting transition. It was shown in the paper[59] that the persistent current is observed in spite of the power dissipation $R_l I_p^2$ since the current decrease because of dissipation is compensated with increasing of the velocity superconducting pairs because of the Bohr quantization at the switching of the loop by fluctuation between superconducting states with different connectivity. Because of these switching the persistent current $I_p \neq 0$ is observed under equilibrium conditions at non-zero dissipation[52-54] just as the equilibrium electric noise measured first by J.B.Johnson[60] and described theoretically by H.Nyquist[61] and is known as Johnson[55] or Nyquist noise. The persistent current $I_p \neq 0$ at $R_l > 0$ can be observed in the fluctuation region near superconducting transition $T \approx T_c$ and the equilibrium current $I_{Ny}$ of the Nyquist noise can be observed in the normal state above $T_c$, for example in the aluminum loop shown on Fig.1. The dissipation of power, $R_l I_p^2$ and $R_l I_{Ny}^2$, is equilibrated in the both cases by the power of thermal fluctuations.

Both the persistent current and the Nyquist noise current are equilibrium phenomenon, but the first, in contrast to the second, is direct current but not random current. Therefore the observation of $I_p \neq 0$ at $R_l > 0$ is experimental evidence of a direct current power observed under equilibrium condition, which can be called persistent power. There is important difference between the persistent power and the equilibrium power of the Nyquist's noise. The power of the Nyquist's noise is "spread" in the frequency region from zero to the quantum limit, whereas the persistent power is not zero at the zero frequency band. Therefore the equilibrium power of the Nyquist's noise is not challenge to the second law of thermodynamics whereas the persistent power is experimental evidence of violation of this fundamental law[52-54]. There is important that not only the dc power $R_l I_p^2$ but also the dc power $V^2/R_{ls}$ is observed on segment of asymmetric superconductor loop[62].

### 3.3 Contradiction between quantization and conservation of momentum circulation
There is other challenge connected with the intrinsic breach of symmetry connected with the Bohr's quantization. The contradiction between quantization and conservation of momentum circulation is clear already from the Meissner effect. This contradiction and its connection with the intrinsic breach of symmetry become more clear when we consider a switching of a loop between superconducting state with different connectivity. Because of the Bohr's quantization (3) the velocity $v_s$ of superconducting pairs in a loop l (like shown on Fig.1 or Fig.2) has discrete spectrum

$$\oint dl v_s = \frac{2\pi\hbar}{m}(n - \frac{\Phi}{\Phi_0}) \qquad (4)$$

in the state with closed wave function, i.e. when the density of superconducting pairs $n_s$ is non-zero along the whole loop. The permitted values of velocity $v_s = (2\pi\hbar/ml)(n - \Phi/\Phi_0)$ is not equal zero at $\Phi \neq n\Phi_0$ and consequently the persistent current $I_p = s2en_s v_s$ should flow along the loop. This current should damp $I(t) = I_p \exp(-t/\tau_{RL})$ during the time of current relaxation $\tau_{RL} = L_l/R_B$ when a segment near the point B is switched at time $t = 0$ in the normal state with the resistance $R_B$. $L_l$ is the inductance of the loop l. Superconducting pair is braked, i.e. its velocity decreases down zero, because of the pure classical electric force $mdv_s/dt = 2eE = -2e\nabla V = -2eV_{-B}/(l-l_B) = 2eV_B/(l-l_B)$, Fig.2, where $V_{-B} = -V_B$ and $V_B = R_B I(t)$ are the potential difference on the segment $l-l_B$ remaining in superconducting state and on the segment switched in the normal state with $R_B > 0$. The pair velocity in any point of the $l-l_B$ segment, for example in the point A on Fig.2, should also change in time $dv_s/dt \neq 0$ when the B segment returns in superconducting state and if $\Phi \neq n\Phi_0$ since the velocity should change from $= 0$ to a permitted value $v_s = (2\pi\hbar/ml)(n - \Phi/\Phi_0)$ at closing of the superconducting state. This acceleration can not be connected with any force. The momentum circulation if superconducting pairs (3) changes from $M_l = q\Phi$ in the state with unclosed wave function to $M_l = n2\pi\hbar$ in the state with closed wave function without any force as well as at the Meissner effect. The forced change of the $M_l$ value $n2\pi\hbar - q\Phi = 2\pi\hbar(n - \Phi/\Phi_0)$ in loop is smaller than at the Meissner effect since $n \neq 0$. But there is important that the velocity in the point A changes because of the change of the state in the point B, without any change in the point A, Fig.2. The calculations, made on base of an obtained experimental results and theory, testify that the persistent current can be observed in enough large loop, for example with the length $l = 10$ m at the section area $s = 1$ μm. There is no reason to doubt that even at this large distance between the points A and B the change of the state in one point can induce the velocity change in other point. It can be challenge to locality when the change occurs because of the Bohr's quantization.

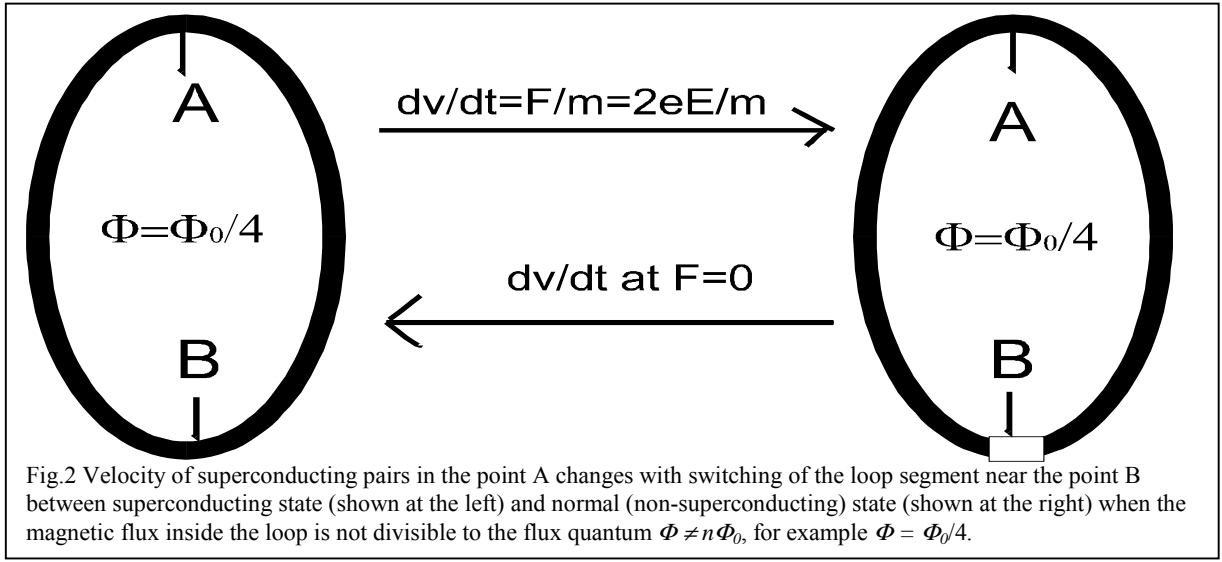

Fig.2 Velocity of superconducting pairs in the point A changes with switching of the loop segment near the point B between superconducting state (shown at the left) and normal (non-superconducting) state (shown at the right) when the magnetic flux inside the loop is not divisible to the flux quantum $\Phi \neq n\Phi_0$, for example $\Phi = \Phi_0/4$.

The dc potential difference $-\langle V_{-B}\rangle = \langle V_B\rangle = L_l\omega_{sw}\langle I_p\rangle$ should be observed both on switched $l_B$ segment and superconducting $l-l_B$ segment when the B segment is switched between superconducting and normal state with a frequency $\omega_{sw} \ll 1/\tau_{RL}$ since the potential difference $V_B(t) = R_B I(t) = R_B I_p \exp(-t/\tau_{RL})$ appears after each transition in the state with $R_B > 0$. The average value of the persistent current $\langle I_p\rangle = s2en_s\langle v_s\rangle = (2\pi\hbar/ml)(\langle n\rangle - \Phi/\Phi_0)$ is not equal zero at $\Phi \neq n\Phi_0$ and $\Phi \neq (n+0.5)\Phi_0$. The average value $\langle n\rangle$ of the quantum number n is close to an integer number corresponding to the minimum energy, i.e. minimum $(n - \Phi/\Phi_0)^2$ since the energy difference between adjacent permitted states is very high, $\gg k_B T$, in real superconducting loop[59]. $\langle n\rangle - \Phi/\Phi_0 = 0$ at $\Phi = (n+0.5)\Phi_0$ when two permitted states $n - \Phi/\Phi_0 = 0.5$ and $n - \Phi/\Phi_0 = -0.5$ have the same minimum energy and opposite directed velocity. Therefore one may expect that the dc potential difference $\langle V_B\rangle = L_l\omega_{sw}\langle I_p\rangle = V(\Phi/\Phi_0) \propto \langle n\rangle - \Phi/\Phi_0$ should be a periodical function of the magnetic flux $\Phi$ inside the loop with the period equals the flux quantum $\Phi_0$ and $V(\Phi/\Phi_0) = 0$ at $\Phi \neq n\Phi_0$ and $\Phi \neq (n+0.5)\Phi_0$. Just such quantum oscillations of the dc voltage $V(\Phi/\Phi_0)$, Fig.1,4, are observed on segment of asymmetric superconducting loops[50,51].

There is important to note that the dc potential difference $V(\Phi/\Phi_0)$ is observed both on switched $l_B$ segment and superconducting $l-l_B$ segment. One could expect that superconducting pairs should accelerate $mdv_s/dt = F_e = 2e\langle E\rangle$ up to infinity (for the critical velocity in the reality) because of the dc electric field $\langle E\rangle = -\nabla\langle V_{-B}\rangle$. But the infinite acceleration does not occur since the change of the velocity from $v_s = (2\pi\hbar/ml)(n - \Phi/\Phi_0)$ to $v_s = 0$ because of $E = -\nabla V_{-B}$ is compensate its change from $v_s = 0$ to $v_s = (2\pi\hbar/ml)(n - \Phi/\Phi_0)$. The latter change of the pair momentum $mv_s$ in a time unity was called in[59] "quantum force". The introducing the "quantum force", equal $F_q = (2\pi\hbar/l)(n - \Phi/\Phi_0)\omega_{sw}$ at a switching frequency $\omega_{sw} \ll 1/\tau_{RL}$, provides formally the force balance: $F_q + 2e\langle E\rangle = 0$. This balance gives the relation between the switching frequency $\omega_{sw}$ and the dc voltage $\langle V_{-B}\rangle = (\pi\hbar\omega_{sw}/e)(n - \Phi/\Phi_0)(1-l_B/l)$ like the Josephson relation[63]. The "quantum force" is not a real force like the electric force $F_e = 2eE$. It does not explain but describes only the change of momentum and momentum circulation because of the Bohr's quantization. The observation of the dc potential difference $\langle V_{-B}\rangle$ on superconducting loop segment is challenge to the law of momentum conservation since the force $F_e = 2e\langle E\rangle$ of the dc electric field $\langle E\rangle = -2e\nabla\langle V_{-B}\rangle$ acting on superconducting pair in the $l-l_B$ segment is not compensated by any real force. This challenge may be connected with the intrinsic breach of right-left symmetry observed in the phenomenon of the quantum oscillation $V(\Phi/\Phi_0)$ of the dc voltage. It is well known that laws of conservation can be substantiated on symmetry demand. There is important that it is impossible to make a switching between states with different connectivity of wave function on atomic level whereas it is possible on mesoscopic level.

**3.4 Principle of impossibility of noninvasive measurement**
Some experts understand that the mesoscopic level may be the field of the collision of quantum mechanics with our immediate experience of the "everyday" words. Most obvious collision is the contradiction between quantum mechanics and macroscopic realism[18]. This problem may be expressed in the question: "Could a Schrodinger's cat exist?" According to the formalism of the quantum mechanics a quantum system can be in a superposition of states but this superposition can not be observed because of its reduction to single state at measuring. Therefore the principle of impossibility of noninvasive measurement is very important in logic of quantum mechanics. The principle of the impossibility of noninvasive measurement seems admissible on the microscopic level when measuring device can not be smaller than measured object. But we can not assume that the Schrodinger cat can die or revive because of our look. The contradiction between quantum mechanics and the possibility of noninvasive measurability[64] may can emerge on the mesoscopic level.

## 4. QUANTUM MECHANICS VERSUS REALISM

The postulate on superposition of quantum states results to contradiction with principle of realism connected not only with the noninvasive measurability. Founders of the quantum mechanics realized this contradiction. The Schrodinger cat manifests the contradiction between quantum mechanics and macroscopic realism. The Einstein-Podolsky-Rosen paradox manifests the contradiction quantum mechanics with local realism. Defending the Copenhagen interpretation Niels Bohr wrote: "*There is no quantum world. There is only an abstract quantum physical description. It is wrong to think that the task of physics is to find out how Nature is. Physics concerns what we can say about Nature*", the citation from[65]. The experimental results obtained in the years after the philosophical discussion between the Founders witness on increased urgency of this discussion[66]. No common consent is about conceptual foundation of quantum physics now as well as in the time of the appearance of this paradoxical science.

**4.1 Entanglement violates the principle of local realism**
It is no coincidence that J. S. Bell called his paper[67] *On the Einstein-Podolsky-Rosen paradox*. Einstein, Podolsky and Rosen in[27] try to prove that the description of reality as given by a wave function is not complete using a paradoxical conclusion from a thought-experiment. They consider quantum systems consisting of two particles which interacted from the time t=0 to t=T, after which time EPR suppose that there is no longer any interaction between the two particles. This supposition by EPR seems very reasonable for the common sense when, for example, the particles are separated by some kilometers or even meters. EPR state also that the objective physical reality should exist with the criterion: *If, without in any way disturbing a system, we can predict with certainty (i.e., with probability equal to unity) the value of a physical quantity, then there exists an element of physical reality corresponding to this physical quantity*. On this basis of the supposition on the local realism EPR have proved that the *wave function does not provide a complete description of the physical reality*.

Indeed, according to the Heisenberg's uncertainty relation $\Delta p \Delta x > \hbar/2$, one of the bases of the Copenhagen interpretation of quantum mechanics, when the momentum of a particle is known $\Delta p = 0$, its coordinate $\Delta x$ has no physical reality. The fundamental principle of the Copenhagen interpretation is the impossibility of noninvasive measurement. We can not measure precisely and simultaneously both momentum and coordinate since any measurement alters the state of quantum particle, a process known as the reduction of the wave function. But because of the law of conservation of momentum the measurement on momentum performed on, say, particle 1 immediately implies for particle 2 a precise momentum even when the two particles are separated by arbitrary distances without any actual interaction between them. Then, if the local realism is valid, i.e. the measurement performed on particle 1 can not alter the state of particle 2, we can define, contrary to the uncertainty relation, precise values both momentum and coordinate particle 2 after the measurement performed on its coordinate. Einstein, Podolsky and Rosen write in the end of the paper that one would not arrive at their conclusion if *the reality of momentum and coordinate of the particle 2 depend upon the process of measurement carried out on the particle 1, which does not disturb the state of the particle 2 in any way*. They state: *No reasonable definition of reality could be expected to permit this*.

Experiments[19-26] have confirmed the quantum mechanical prediction contrary to the local realism. These investigations stimulated the appearance of the idea of quantum computation based on the entanglement, paradoxical quantum phenomenon violating the principle of the local realism. Therefore a possibility of quantum computation is connected with violation of the principles of realism. In order to find this violation in superconductor structures the contradiction of quantum mechanics with realism ought be realized and investigated.

**4.2 Contradiction between results of measurements following from principle of quantum mechanics**
Measurements play a radically different role in the quantum world than they do in the classical world. Classically, measurements have an essentially passive nature: they find out about a pre-existing reality. But an assumption on a pre-existing reality contradicts to the bases of the Copenhagen interpretation of the quantum mechanics. This contradiction can bring to a situation when measurements could give mutually contradictory results.

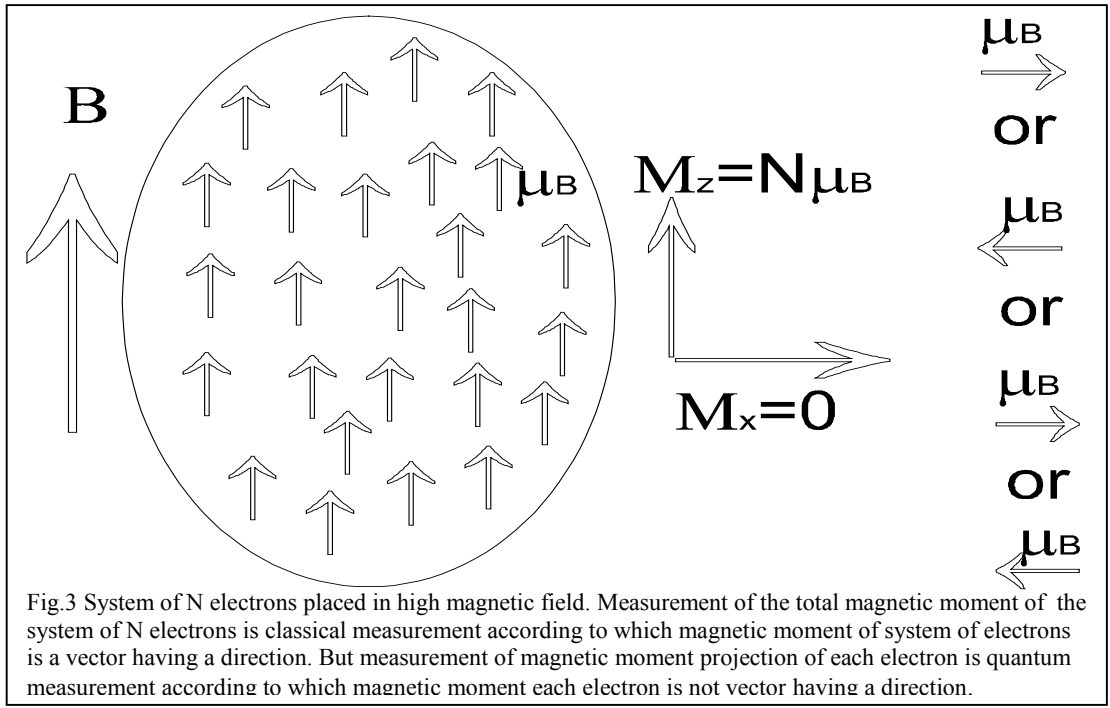

Fig.3 System of N electrons placed in high magnetic field. Measurement of the total magnetic moment of the system of N electrons is classical measurement according to which magnetic moment of system of electrons is a vector having a direction. But measurement of magnetic moment projection of each electron is quantum measurement according to which magnetic moment each electron is not vector having a direction.

It can be, for example, at measurements of magnetic moment of a system great numbers N of elementary particles (electrons) with spin ½. According to the principles of quantum mechanics spin has not a direction and the spin magnetic moment of electron and projection of the magnetic moment measured along any axis equal $\mu_B = -e\hbar/2m$. Measurement of magnetic moment of the system of N electrons placed in very high magnetic field $B = B_z \gg k_B T/\mu_B$ along the z-axis, i.e. along the magnetic field, should give value $M_z = N\mu_B$ and measurement along x should give zero value $M_x = 0$, Fig.3. This means that the projection on z-axis of spin magnetic moment of each electron equals $\mu_B$. But according to the principle of quantum mechanics measurement of spin magnetic moment of each electron along x-axis should give value $\mu_B$ or $-\mu_B$, Fig.3. There is obvious contradiction between results these measurements since spin magnetic moment of electron equals $\mu_B$ but not $\sqrt{2}\mu_B$. We can conclude that according to quantum mechanics and contrary of classical logic the contradiction between results of two measurements does not mean that even if one measurement is not correct.

**4.3 Quantum mechanics versus macroscopic realism**
We do not wonder already that a quantum system can be at the same time in several states but it is funny that enough many authors have no doubt that the Schrodinger's cat can die or revive because of our look and declare on experimental evidence for a coherent superposition of macroscopically distinct quantum states[68-70], although the validity of the principle of impossibility of noninvasive measurement raises doubts on mesoscopic and higher levels. These authors disregard the question "Is the flux there when nobody looks?"[18] or "Is the moon there when nobody looks?"[71] which should be raised if a superposition of macroscopic quantum states can indeed exist.

The experts in quantum physics, following to Schrodinger, raised these questions in order to fix our thoughts on the paradoxical nature of quantum theory. Although the Schrodinger's cat paradox has exerted a perpetual fascination over many physicists since its publication in 1935, it is fair to say that not all physicists understand this paradox. A scholastic approach to the interpretation of this paradox predominates now. Many authors try to describe the unhappy cat with help of a wave function[66]. Moreover, some physicists say about entanglement between atom and cat states, although it is obvious from the consideration by Schrodinger that the connection of the cat state with the atom state, by means of a Geiger counter and a small flask of hydrocyanic acid, is pure classical.

The question "Is the flux there when nobody looks?" in[18] concerns the magnetic flux $L_l I_p$ induced by the persistent current $I_p$ in superconducting loop. Two permitted states n and n+1 have the same minimum energy at the magnetic flux inside the loop $\Phi = (n+0.5)\Phi_0$. It is possible to measure the persistent current magnetic flux $L_l I_p$ or $-L_l I_p$ in each of these states. But what could be if the superposition of these states exists? It is obvious that we can not see or measure the superposition. According to the principle of quantum mechanics we can see one or other permitted state with the magnetic flux $L_l I_p$ or $-L_l I_p$. But in which moment of measurement could the superposition of states reduce to single state? Quantum mechanics can not answer on this question. Quantum mechanics can not describe the process of quantum measurement[66]. Therefore one ought rely only on experiment at the investigation of a possibility the quantum superposition of macroscopic state in a superconducting loop.

### 4.4 Contradiction between results of measurements observed on mesoscopic level
One should first of all makes sure on base of reliable experimental results that two permitted states indeed exists in superconducting loop at $\Phi = (n+0.5)\Phi_0$. After that one should makes sure that multiple measurements give a result corresponding an average value between two states whereas single measurement gives a result corresponding to one of the two permitted states.

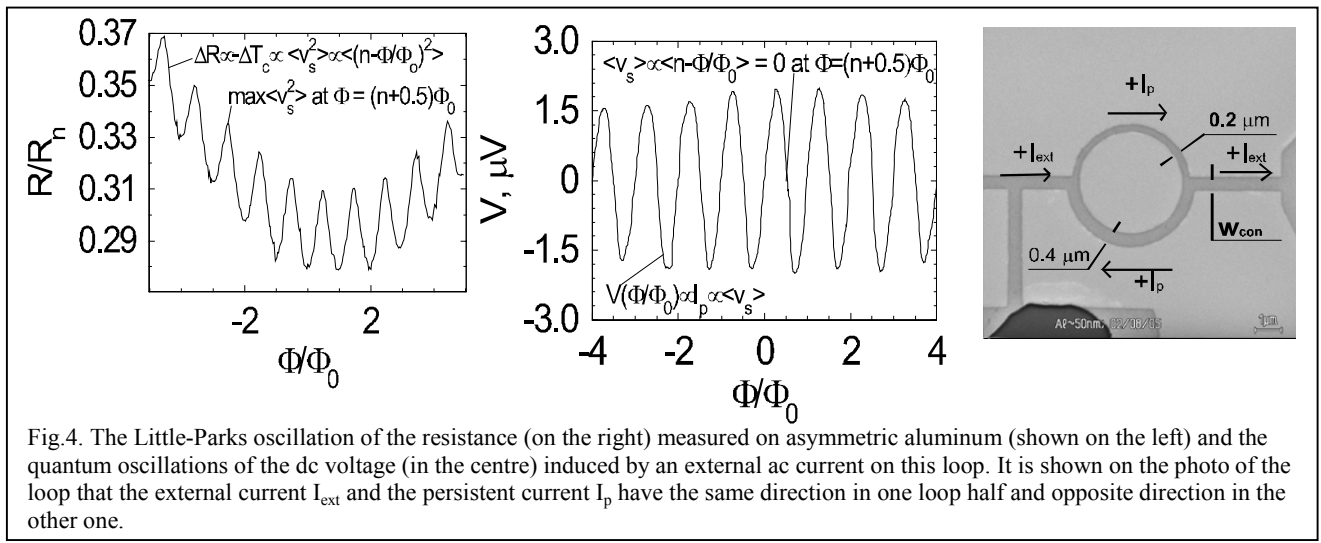

Fig.4. The Little-Parks oscillation of the resistance (on the right) measured on asymmetric aluminum (shown on the left) and the quantum oscillations of the dc voltage (in the centre) induced by an external ac current on this loop. It is shown on the photo of the loop that the external current $I_{ext}$ and the persistent current $I_p$ have the same direction in one loop half and opposite direction in the other one.

There are experimental results which can be considered as experimental evidence of two permitted states in superconducting loop. The measurements of the Little-Parks oscillations $R(\Phi/\Phi_0)$ of the resistance and the quantum oscillation of the dc voltage $V(\Phi/\Phi_0)$ induced by an external ac current made on the same asymmetric loop have shown that maximums of the resistance are observed at $\Phi = (n+0.5)\Phi_0$ whereas the rectification voltage equals zero at these values of the magnetic flux inside the loop. The Little-Parks oscillations $R(\Phi/\Phi_0)$ are observed since the energy $\Delta E \propto v_s^2 \propto (n - \Phi/\Phi_0)^2$ of the closed superconducting state increases because of non-zero pair velocity $v_s \propto (n - \Phi/\Phi_0) \neq 0$ at $\Phi \neq n\Phi_0$, when the state with zero velocity is forbidden[36]. The energy increase results in the decrease of the critical temperature $-\Delta T_c \propto \Delta E \propto v_s^2$ and as consequence to the increase of the resistance $\Delta R \propto -\Delta T_c \propto \Delta E \propto v_s^2$ at the temperature of measurements. The oscillations $R(\Phi/\Phi_0)$ are measured near superconducting transition, at $0 < R < R_n$, where thermal fluctuations switch the loop between superconducting state with different connectivity. Therefore these measurements correspond to average value of the velocity square $\Delta R(\Phi/\Phi_0) \propto <v_s^2> \propto <(n - \Phi/\Phi_0)^2>$.

The measurement of the rectification voltage $V(\Phi/\Phi_0)$ corresponds to multiple measurements of the persistent current $V(\Phi/\Phi_0) \propto \langle I_p \rangle \propto \langle v_s \rangle \propto \langle n - \Phi/\Phi_0 \rangle$. These two measurements show that the average value of the velocity square $\langle v_s^2 \rangle \propto \langle (n - \Phi/\Phi_0)^2 \rangle$ has maximum value at $\Phi = (n+0.5)\Phi_0$ whereas the average velocity equals zero $\langle v_s \rangle \propto \langle n - \Phi/\Phi_0 \rangle = 0$. These results are experimental evidence that there are indeed two permitted states at $\Phi = (n+0.5)\Phi_0$ with $v_s \propto n - \Phi/\Phi_0 = \frac{1}{2}$ and $v_s \propto n - \Phi/\Phi_0 = -\frac{1}{2}$: $\langle v_s^2 \rangle \propto \langle (n - \Phi/\Phi_0)^2 \rangle \propto (\frac{1}{2})^2 + (-\frac{1}{2})^2$ has maximum value whereas $\langle v_s \rangle \propto \langle n - \Phi/\Phi_0 \rangle \propto \frac{1}{2} + (-\frac{1}{2}) = 0$.

Recently measurements[72] were made which corresponding single measurement of the quantum state of superconducting loop. The quantum oscillations of the dc voltage $V(\Phi/\Phi_0)$ induced by an external as current $I_{ext}(t) = I_0 \sin(2\pi ft)$ may be interpreted as a result of rectification because of asymmetry of the current-voltage curves[51]. This asymmetry takes place and its value and sign are periodical function of magnetic because of superposition of the external and persistent currents in the narrow loop half $j_n = I_{ext}/(s_n+s_w) \pm I_p/s_n$ and the wide loop half $j_w = I_{ext}/(s_n+s_w) \pm I_p/s_w$. The current density mounts the critical value in the both half $j_n = j_w = I_{ext}/(s_n+s_w) = j_c$ at $|I_{ext}|_c = (s_n+s_w)j_c$ when the persistent current $I_p = 0$ at $\Phi = n\Phi_0$. The persistent current increases the current density in narrow $j_n$ or wide $j_w$ half. Therefore the external critical current measured in the left-right direction should be equal $|I_{ext}|_{c+} = (s_n+s_w)(j_c - I_p/s_n)$ at the clockwise direction of $I_p$, see Fig.4, and $|I_{ext}|_{c+} = (s_n+s_w)(j_c - |I_p|/s_w)$ at the clockwise direction of $I_p$. The critical current measured in the right-left direction should be equal $|I_{ext}|_{c-} = (s_n+s_w)(j_c - I_p/s_w)$ at the clockwise direction of $I_p$, see Fig.4, and $|I_{ext}|_{c+} = (s_n+s_w)(j_c - |I_p|/s_n)$ at the clockwise direction of $I_p$. The value and direction of the persistent current can be obtained from measurements of the critical current of asymmetric loop since $|I_{ext}|_{c-} - |I_{ext}|_{c+} = I_p(s_w/s_n - s_n/s_w)$, where positive value of $I_p$ corresponds the clockwise direction, Fig.4. Measurements give periodical dependence $|I_{ext}|_{c+}(\Phi/\Phi_0)$, $|I_{ext}|_{c-}(\Phi/\Phi_0)$ of aluminum asymmetric loop shown on Fig. 4, with the asymmetry of the critical current $|I_{ext}|_{c-} - |I_{ext}|_{c+} \neq 0$ at $\Phi \neq n\Phi_0$ and $\Phi \neq (n+0.5)\Phi_0$. The measured dependence $|I_{ext}|_{c-}(\Phi/\Phi_0) - |I_{ext}|_{c+}(\Phi/\Phi_0) \propto \langle n - \Phi/\Phi_0 \rangle$ is like $V(\Phi/\Phi_0)$, see Fig.4, i.e. corresponds to multiple measurement but no single measurement as one should expect. Moreover, the minimums of the $|I_{ext}|_{c+}(\Phi/\Phi_0)$ and $|I_{ext}|_{c-}(\Phi/\Phi_0)$ are observed at $\Phi = (n+0.25)\Phi_0$ and $\Phi = (n+0.75)\Phi_0$ but no at $\Phi = (n+0.5)\Phi_0$ as one should be expect. These results of measurements of $|I_{ext}|_{c+}(\Phi/\Phi_0)$ and $|I_{ext}|_{c-}(\Phi/\Phi_0)$ come into irreconcilable contradictions with results of measurements of the Little-Parks oscillations $\Delta R(\Phi/\Phi_0)$ obtained on the same aluminum loop, Fig.4. According to the obtained $\Delta R(\Phi/\Phi_0)$ dependence the persistent current has maximum value at $\Phi = (n+0.5)\Phi_0$, whereas the measurements of the critical current dependencies $|I_{ext}|_{c+}(\Phi/\Phi_0)$ and $|I_{ext}|_{c-}(\Phi/\Phi_0)$ give evidence that the position of these maximums depends on the direction of the measuring current and differs from $\Phi = (n+0.5)\Phi_0$. This contradiction may be a manifestation of the conflict of quantum mechanics with the principle of realism observed on mesoscopic level.

## 5. POSSIPLE EXPERIMALTAL VERIFICATION OF THE EPR CORRELATION IN SUPERCONDUCTOR STRUCTURES

The paradoxical consequences of quantum principles observed on the mesoscopic and macroscopic levels show on the one hand the fundamental difference between these levels and atomic level and on the other hand they give a hope of a possibility of the entanglement on the levels higher than atomic. In this Section we propose an experimental test of the entanglement possibility in superconductor structure. The EPR paradox[27] and the Bell's experiments[19-26] testify that the entanglement is a combination of a law of conservation and quantum superposition. Therefore our proposals to verify the possibility of entanglement are based on the law of conservation of momentum circulation of superconducting pairs.

We know from numerous experimental results that the equilibrium persistent current measured in a superconducting loop with week screening $L_l I_p \ll \Phi_0$ changes periodically with magnetic field $I_p(\Phi/\Phi_0) \propto \langle n \rangle - \Phi/\Phi_0$. This means that the momentum circulation of superconducting pairs $M_l = n2\pi\hbar$ changes at some values of the magnetic flux inside the loop. Superconducting state in the three loops united by superconducting stripes (shown in the top of Fig.5) are described by common wave function. We can foreknow that the entanglement is possible in this case and then a correlation can be expected between the values of the momentum circulation $M_{l,l} = n_l 2\pi\hbar$ $M_{l,m} = n_m 2\pi\hbar$ and $M_{l,r} = n_r 2\pi\hbar$ in loops.

This correlation can be detected by comparison of the measurements of the magnetic dependencies of the persistent current in the middle loop when it is united and is not united with two other loops by superconducting stripes, Fig.5. The middle loop may be made with two or more Josephson junction, like the flux qubit.

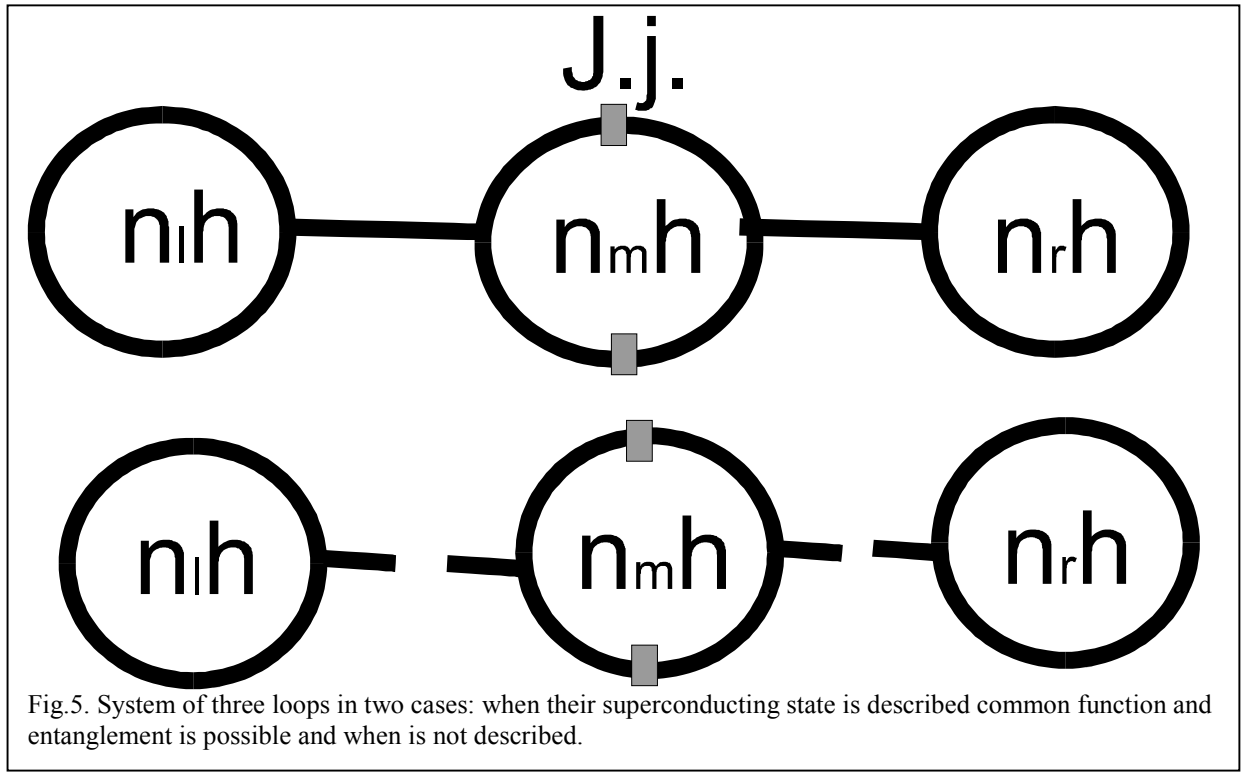

Fig.5. System of three loops in two cases: when their superconducting state is described common function and entanglement is possible and when is not described.

**ACKNOWLEDGMENTS**

This work was financially supported by ITCS department of Russian Academy of Sciences in the Program "Technology Basis of New Computing Methods", by Russian Foundation of Basic Research (Grant 04-02-17068) and by the Presidium of Russian Academy of Sciences in the Program "Low-Dimensional Quantum Structures".